\documentclass[twocolumn,showpacs,preprintnumbers,amsmath,amssymb]{revtex4}
\usepackage{graphicx}% Include figure files
\usepackage{dcolumn}% Align table columns on decimal point
\usepackage{bm}% bold math

\newcommand{\opr}[1]{\ensuremath{\mathbf{\mathsf{#1}}}}

\newcommand{\kernel}[1]{\ensuremath{\left<q\left|#1\right|q'\right>}}

\newcommand{\ket}[1]{\ensuremath{\left|#1\!\left.\right>\right.}}

\begin{document}
%\title{On Decoherence and the Quantum Measurement Problem}% Force line breaks with \\
%\author{Roland Cristopher F. Caballar}
% \email{roland_cristopher.caballar@up.edu.ph}
%\affiliation{%
%Theoretical Physics Group, National Institute of Physics, University of the Philippinies, Diliman, Quezon City 1101}%
\date{\today}% It is always \today, today,
\title{Characterizing Multiple Solutions to the Time-Energy Canonical Commutation Relation via Quantum Dynamics}


%\author{Roland Cristopher F. Caballar}
%\ead{rfcaballar@up.edu.ph}
%\author{Eric A. Galapon}
%\ead{eric.galapon@up.edu.ph}
%\address{Theoretical Physics Group, National Institute of Physics\\University of the Philippinies, Diliman, Quezon City 1101}
\author{Roland Cristopher F. Caballar}
\email{rcaballar@nip.upd.edu.ph}
\author{Eric A. Galapon}
\email{eric.galapon@up.edu.ph}
\affiliation{Theoretical Physics Group, National Institute of Physics\\University of the Philippinies, Diliman, Quezon City 1101}
\begin{abstract}
We address the multiplicity of solutions to the time-energy canonical commutation relation for a given Hamiltonian. Specifically, we consider a particle spatially confined in a potential free interval, where it is known that two distinct self-adjoint and compact time operators conjugate to the system Hamiltonian exist. The dynamics of the eigenvectors of these operators indicate that different time operators posses distinguishing properties that can unambiguously associate them to specific aspects of the quantum time problem.
\end{abstract}

%\begin{keyword}
%Time Operator\sep Transition Probability\sep Quantum Dynamics\sep Canonical Commutation Relations
%\pacs{03.65.Ta, 02.70.Hm}
%\PACS 03.65.Ta\sep 02.70.Hm
%\end{keyword}
\maketitle
% main text
\section{Introduction}
Quantum mechanics' role as a fundamental theory of modern physics is illustrated by the fact that Newtonian mechanics is regarded as a limiting case of quantum mechanics. However, discrepancies still exist between Newtonian and quantum mechanics that will weaken any correspondence between them. One of these discrepancies concerns the role of time as a dynamical observable in quantum mechanics. Unlike in Newtonian mechanics, where time's role as a dynamical observable is well-defined, for example, time of arrival is expressed as a function of the state variables position and momentum, the exact nature of time as a dynamical observable in quantum mechanics still remains to be resolved \cite{muga3}. The problem of determining time's exact role as an observable in quantum mechanics, known as the quantum time problem, has many aspects, one of which involves determining if there exist self-adjoint and Hamiltonian conjugate time operators representing dynamical time observables. Until just recently the answer is in the negative---no self-adjoint and conjugate time operator exists for semibounded Hamiltonians \citep{pauli,gian,holevo1,park,srinivas,olhovsky,toller2,holland,nuemann1,sakurai,gottfried,omnes,blanchard,harald,busch1,allcock,kijowski,muga2}. This belief has been precipitated by Pauli's well-known non-existence theorem (PT) for such operators \cite{pauli}. 

However, it is now known that a counter example to Pauli's theorem exists. The first has been given by one of us in the form of the quantized free time of arrival expression for a spatially confined particle under certain boundary conditions \citep{phd,gal021}. The counter example is a class of bounded, compact and self-adjoint operators conjugate to a semibounded and discrete Hamiltonian; these are now known as the confined time of arrival operators \citep{gal04,gal05,gal06a}. It will later become clear that not only such operators exist but that a given Hamiltonian maybe conjugate to several non-unitary equivalent self-adjoint time operators \cite{gal06}. For certain Hamiltonians, these self-adjoint time operators are compact so that they posses a bounded, discrete spectrum. However, their compactness presents interpretational problems: What meaning can be attached to such non-covariant operators? How such operators fit into the currently accepted scheme of quantum measurement theory? For the confined time of arrival operators, it is determined that their proper interpretation hinges on the dynamics of the eigenfunctions---the evolving eigenfunctions assuming specific properties at their respective eigenvalues \citep{gal04,gal05,gal06a}. This unequivocally demonstrates that such time operators can be meaningful.

In this paper we address the multiplicity of solutions to the time-energy canonical commutation relation for a given Hamiltonian. Specifically, we consider a particle spatially confined in a potential free interval. It is known that two distinct self-adjoint and compact time operators correspond to the Hamiltonian of the system \cite{gal06}. The overriding question is whether these two time operators can be ascribed distinct physical interpretations. This is important because the existence of multiple time operators corresponding to a given Hamiltonian raises the plausibility of associating the different facets of the quantum time problem---quantum time of arrival, quantum traversal time, quantum tunneling time---to the different time operators of the system. Indeed we will find, at least for the system at hand, that the different time operators posses distinguishing properties that can unambiguously associate them to specific aspects of the quantum time problem.

\section{Multiple Solutions to the The Time - Energy Canonical Commutation Relation}
First let us examine the time-energy canonical commutation relation (TE-CCR). Given the Hamiltonian $\opr{H}$, with domain $\mathcal{D}_{\opr{H}}$ in the system Hilbert space $\mathcal{H}$, we refer to an operator $\opr{T}$, with domain $\mathcal{D}_{\opr{T}}$ in $\mathcal{H}$, as a time operator or a solution to the time-energy canonical commutation relation if there exists a non-trivial subspace $\mathcal{D}_c$ of the Hilbert space $\mathcal{H}$ consisting of elements of both the domains of the Hamiltonian $\opr{H}$ and the operator $\opr{T}$ such that for all $\ket{\varphi}$ in $\mathcal{D}_c$ the canonical commutation relation $[\opr{T},\opr{H}]\ket{\varphi}=i\hbar\ket{\varphi}$ or $[\opr{H},\opr{T}]\ket{\varphi}=i\hbar\ket{\varphi}$ holds. A time operator, together with its Hamiltonian, can be characterized as to whether the canonical domain $\mathcal{D}_c$ is dense or not; in \cite{gal06} we have referred to a pair as of dense category solution to the CCR if the the canonical domain is dense, otherwise, of closed category. 

Now if the relation $[\opr{T},\opr{H}]\ket{\varphi}=i\hbar\ket{\varphi}$ holds, we refer to $\opr{T}$ as a passage-time-type (PTT) time operator; if instead the relation $[\opr{H},\opr{T}]\ket{\varphi}=i\hbar\ket{\varphi}$ holds, we refer to $\opr{T}$ as a time-of-arrival-type (TAT) time operator. These designations of the solutions to the TE-CCR are based on the explicit forms of the Poisson brackets which the classical passage time and classical arrival time obey. In particular, in classical mechanics, the explicit form of the Poisson bracket for the classical arrival time, which holds only at $t=t_{0}$, is $\left\{H,T_{A}\right\}=1$ where, say, for a free particle with mass $\mu$ and position and momentum $q_0$ and $p_0$ at the initial instant of time $t=t_0$, respectively, the classical time of arrival is $T_{A}=-\mu q_{0}p_{0}^{-1}$. On the other hand, the explicit form of the Poisson bracket for the classical passage time, which holds for all times $t$, is $\left\{H,T_{P}\right\}=-1$ where, say, for a free particle with mass $\mu$ and position and momentum $q$ and $p$, taken at an arbitrary time $t$, $T_{P}=\mu qp^{-1}$ is the classical passage time. Note, however, that even though our definitions for the PTT and TAT time operators were inspired by the classical passage time and arrival time, respectively, the exact nature of the relationship between the TE-CCR and the Poisson brackets for the classical passage and arrival time is not yet clear and is beyond the scope of this paper, since the exact nature and role of time in quantum mechanics is not yet clear to date. As a matter of fact, we note that the TE-CCRs considered in this manuscript were constructed without reference to the Poisson brackets. As such, one must bear in mind that the designation of the solutions to the TE-CCR of the form $[\opr{T},\opr{H}]\ket{\varphi}=i\hbar\ket{\varphi}$ and $[\opr{T},\opr{H}]\ket{\varphi}=-i\hbar\ket{\varphi}$ as PTT and TAT time operators do not by any means signify that these explicit forms of the TE-CCR are derived from the Poisson brackets corresponding to the classical passage and arrival times, respectively. 

Now the difference between a PTT-time operator and a TAT-time operator appears trivial because based on the explicit forms of the TE-CCR which TAT and PTT time operators follow, one can apparently transform one to another by a mere change in sign. A question then is whether one can really discriminate one time operator solution from its negative and from other distinct time operator solutions. That is, is there really a time operator that is intrinsically a PTT or a TAT? Moreover, are there identifying differences between solutions belonging to different categories? We will address these questions by investigating the dynamical behaviors of the eigenvectors of the time operator solutions to the TE-CCR.

\section{Physical System in Which the TE-CCR is Defined}

Let us consider, as the physical system in which we define the TE-CCR, a structureless particle with mass $\mu$ constrained to move one-dimensionally in a potential-free segment $[-l,l]$ \citep{gal04,gal05}. We attach the Hilbert space $\mathcal{H}=L^{2}\left[-l,l\right]$ to the system. The Hamiltonian corresponding to the system is assumed to be conservative and is required to be purely kinetic. The former condition requires the Hamiltonian to be self-adjoint and the latter requires the momentum operator to be self-adjoint as well. The self-adjointness of the momentum operator imposes non-vanishing boundary conditions at $\pm l$. This leads to the ring of momentum operators $(\opr{p}_{\gamma}\varphi)\!(q)=-i\hbar\varphi'(q)$, where $0\leq|\gamma|\leq\pi/2$, defined on all $\varphi(q)$ in $\mathcal{H}$ with square integrable first derivatives and satisfying the boundary condition $\varphi(-l)=e^{-2i\gamma}\varphi(l)$. Then, for a given $\gamma$, the Hamiltonian of the system is given by $\opr{H}_{\gamma}=\opr{p}_{\gamma}^{2}/2\mu$, which is self-adjoint and purely kinetic. The momentum and Hamiltonian operators will then commute, and the spectrum of the Hamiltonian is discrete and semibounded from below. The Hamiltonian and momentum operators will then have as their eigenfunctions $\varphi_{k}^{\gamma}(q)=(2l)^{-1/2}e^{i(\gamma+k\pi)\frac{q}{l}}$, while the eigenvalues of the momentum and Hamiltonian operators are $p_{\gamma,k}=\hbar(\gamma+k\pi)/l$ and $E_{\gamma,k}=p_{k,\gamma}^{2}/2\mu$ respectively. In this paper we consider only the non-degenerate case $\gamma\neq 0,\pi/2$.

\section{A Solution to the TE-CCR of Closed Category}

\subsection{The CTOA Operator}

One time operator that can be constructed for the system is the confined time of arrival (CTOA) operator \citep{gal04,gal05,gal06}. The CTOA-operator can be derived by symmetrically quantizing the classical time of arrival  about the origin, $t=-\mu q/p$ , with $q$ and $p$ replaced by the position and momentum operators, $\opr{q}$ and $\opr{p}_{\gamma}$, of the confined particle, yielding $\opr{T}_{1,\gamma}=-\mu(\opr{q}\opr{p}_{\gamma}+\opr{p}_{\gamma}\opr{q})/2$. In position representation, $\opr{T}_{1,\gamma}$ is given by the integral operator $\left(\opr{T}_{1,\gamma}\varphi\right)\!(q)=\int_{-l}^{l}\kernel{\opr{T}_{1,\gamma}}\varphi(q') dq'$, whose kernel %$\kernel{\opr{T}_{1,\gamma}}$ 
is given by
\begin{eqnarray}
\kernel{\opr{T}_{1,\gamma}}=-\frac{\mu}{4\hbar\sin\gamma}(q+q')\!\left[e^{i\gamma}\mathrm{H}(q-q')\!+\!e^{-i\gamma}\mathrm{H}(q'-q)\right].\nonumber
\end{eqnarray}
The kernel is symmetric and square integrable in the plane $[-l,l]\times[-l,l]$. Hence, the CTOA operator is compact and self-adjoint. Furthermore, the CTOA operator is, by construction, a TAT-type solution to the TE-CCR, with canonical domain $\mathcal{D}_{c,1}$ consisting of elements $\varphi(q)$ in the domain of $\opr{H}_{\gamma}$ satisfying all of the following conditions: $\int_{-l}^{l}\varphi(q')dq'=0,\varphi\left(\pm l\right)=0,\varphi'\left(\pm l\right)=0$. The canonical domain $\mathcal{D}_{c,1}$ is not dense because the vector $\psi(q)=constant$ is orthogonal to all vectors of $\mathcal{D}_{c,1}$. Hence, the CTOA operator is a time operator of closed category.

The compactness of $\opr{T}_{1,\gamma}$ implies that it possesses a complete set of square integrable eigenfunctions with a corresponding discrete spectrum. The unnormalized eigenfunctions are given by
\begin{eqnarray}
\varphi_{n,\gamma}^{\pm}(q)&=&e^{\mp ir_n\frac{q^2}{l^2 }}\left[ J_{\frac{3}{4},\frac{1}{4}}^{\mp}\left( r_n\frac{q^2}{l^2}\right)\!\left(J_{-\frac{1}{4}}(r_n)-\cot\gamma J_{\frac{3}{4}}(r_n)\right) \right.\nonumber\\
& & \left. 
\pm \frac{2 q \sqrt{r_n}}{l} J_{\frac{1}{4},\frac{3}{4}}^{\mp}\left( r_n\frac{q^2}{l^2}\right)\! \left({J_{-\frac{3}{4}}(r_n)-\cot\gamma J_{\frac{1}{4}}(r_n)}\right)\right],\nonumber
\end{eqnarray}
where $J_{\nu,\rho}^{\mp}(x)=(4x)^{\nu}(J_{-\nu}(x)\mp i J_{\rho}(x))$ and the $r_n$'s are the positive solutions of $J_{-3/4}(x) J_{-1/4}(x) -\cot^2\gamma J_{3/4}(x)J_{1/4}(x)=0$, with the sign indicating the sign of the corresponding eigenvalue. The corresponding eigenvalues are given by $\tau_{n,\gamma}^{\pm}=\pm \mu l^2/4\hbar r_n$. (We refer the reader to \citep{gal04,gal05} for a detailed discussion of the CTOA-operator eigenvalue problem.)

\subsection{Physical Interpretation of the CTOA Operator}

As was shown via numerical simulations in references \citep{gal04,gal05}, the CTOA-eigenfunctions evolve according to Schrodinger's equation such that the {\it event of the position expectation value assuming the arrival point}, and the {\it event of the position uncertainty being minimum} occur at the same instant of time equal to their corresponding eigenvalues. Figure 1 shows the behavior of an evolving CTOA-operator eigenfunction, together with the variance of the position operator with time. We have referred to this dynamical behavior of the CTOA-eigenfunctions as unitary arrival at the given arrival point \citep{gal04,gal05}. This observed dynamical behaviors of the eigenfunctions  identifies the CTOA-operators as legitimate time of arrival operators, in fact, first time of arrival operators because the eigenvalues are the first times of unitary arrival of the eigenfunctions at the arrival point. As such, one can physically interpret the CTOA operator as an operator whose time - evolved eigenfunctions unitarily arrive at the origin at an instant of time equal to their corresponding eigenvalues. Such an interpretation can be used to formulate a theory of quantum wavefunction collapse on the appearance of a particle, which is detailed in reference \cite{gal08}.

%%%%%%%%%%%%%%%%%%%%%%%%%%%%%%%%%%%%%%%%%%%%%%%%%%%%%%%%%%%%%%%%%%%%%%%%%%%%%%%%%%%%%%
%%%%%%%%%%%%%%%%%%%%%%%%%%%%%%%%%%%%%%%%%%%%%%%%%%%%%%%%%%%%%%%%%%%%%%%%%%%%%%%%%%%%%%
\begin{figure}
\includegraphics[width=0.23\textwidth,height=0.18\textwidth]{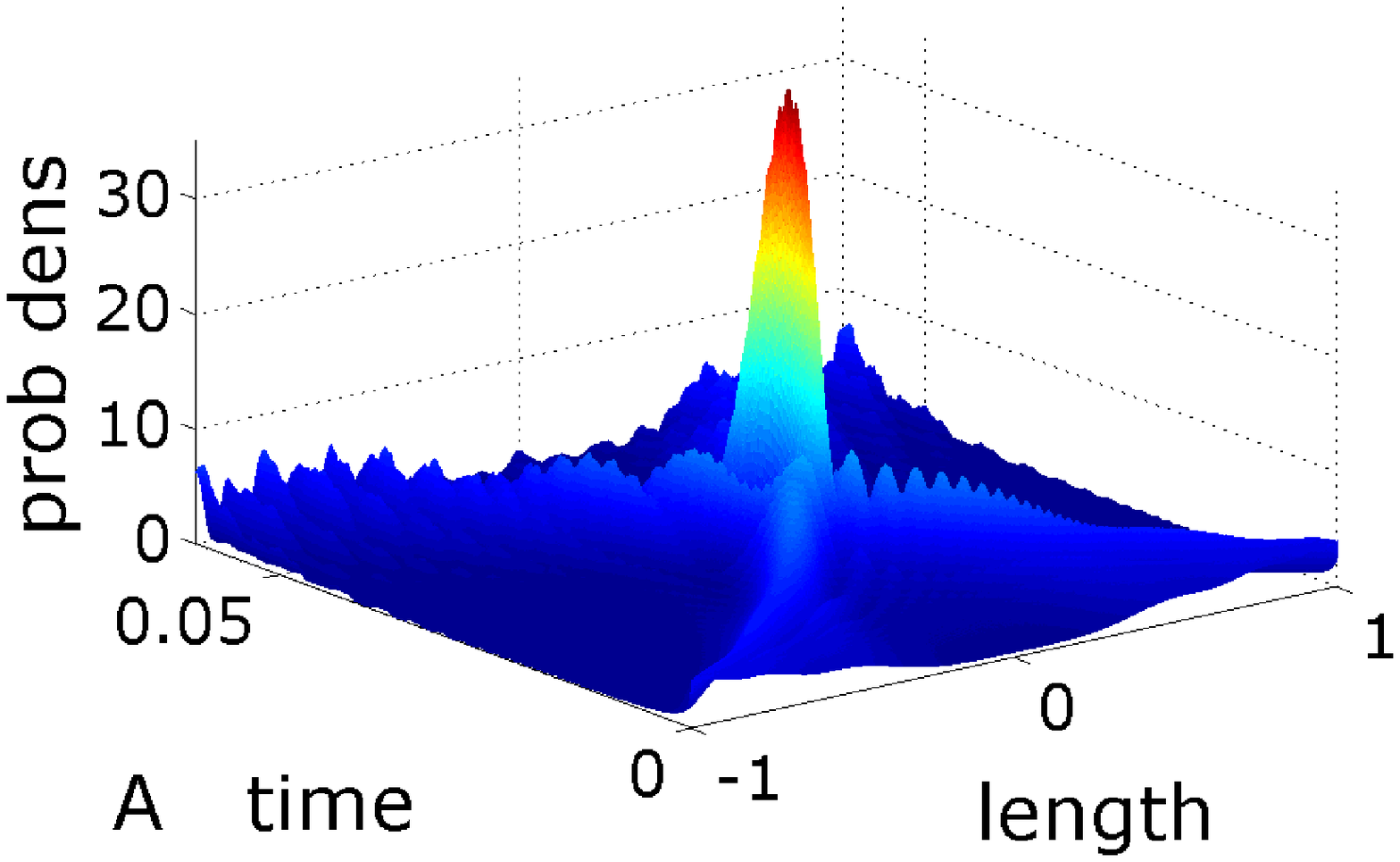}
\includegraphics[width=0.23\textwidth,height=0.18\textwidth]{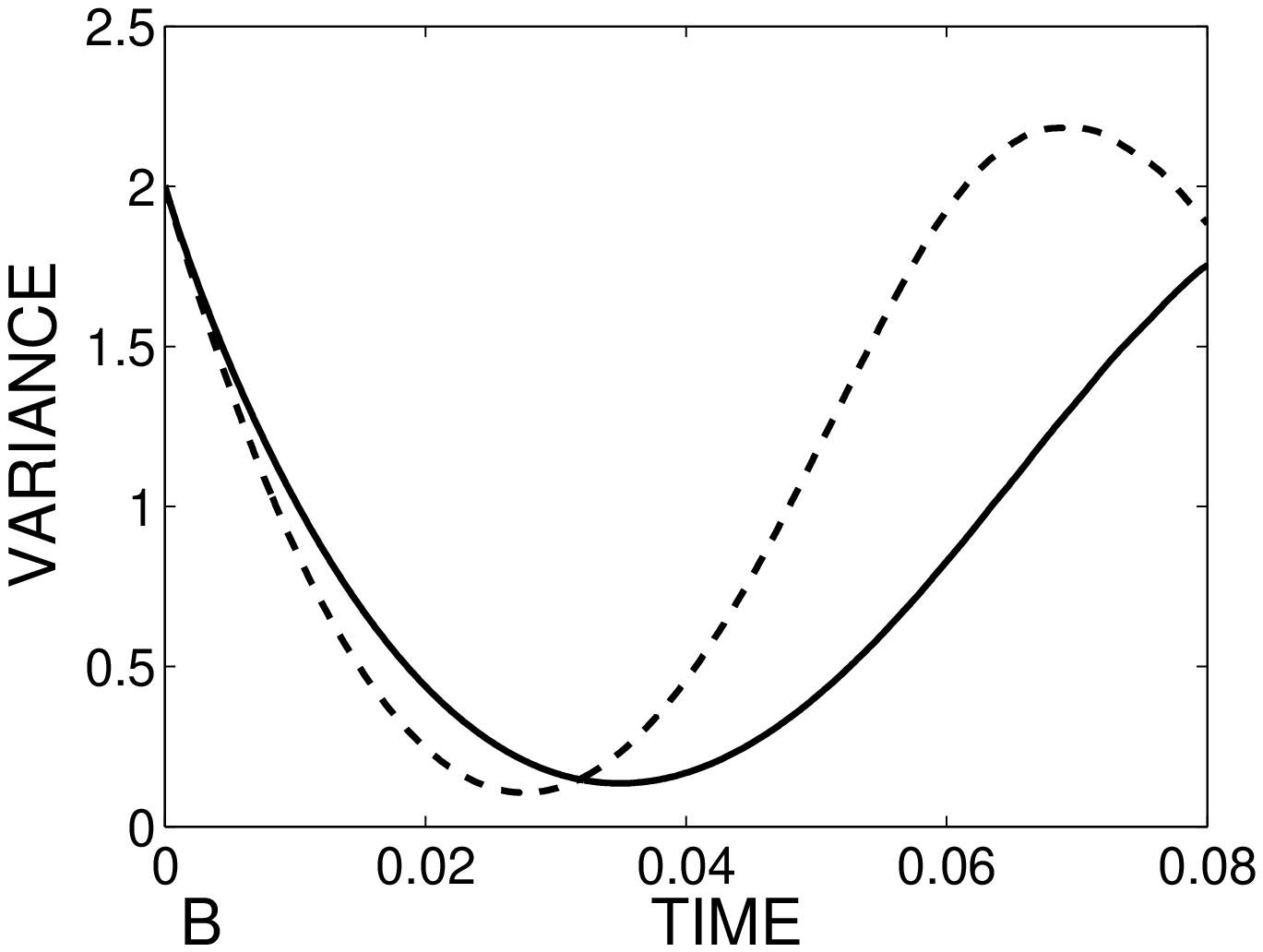}
\caption{(A) Plot of the position probability density of the forward time evolved CTOA eigenfunction corresponding to eigenvalue $\tau_{n,\gamma}=0.02765$ and $\gamma=0.01$. (B) Plot of the variance of the evolving CTOA eigenfunctions corresponding to eigenvalues $\tau_{n,\gamma}=0.02765$ (dotted line) and $\tau_{n,\gamma}=0.03758$ (solid line), with $\gamma=0.01$. All units, as in all succeeding figures, are specified in a.u., with $l=\mu=\hbar=1$.}
\end{figure}
%%%%%%%%%%%%%%%%%%%%%%%%%%%%%%%%%%%%%%%%%%%%%%%%%%%%%%%%%%%%%%%%%%%%%%%%%%%%%%%%%%%%%%
%%%%%%%%%%%%%%%%%%%%%%%%%%%%%%%%%%%%%%%%%%%%%%%%%%%%%%%%%%%%%%%%%%%%%%%%%%%%%%%%%%%%%%

Having provided a physical interpretation for the CTOA operator in terms of the system's internal unitary dynamics, we now ask the following question: is it possible to interpret the CTOA-operators independent of the concept of time of arrival? The following solution to the TE-CCR will give us an opportunity to answer this question.

\section{A Solution to the TE-CCR of Dense Category}

\subsection{The Characteristic Time Operator}

Now another time operator that can be constructed for the same system is the integral operator $(\opr{T_{2,\gamma}}\psi)\!(q)=\int_{-l}^l\kernel{\opr{T}_{2,\gamma}}\psi(q') dq'$, whose kernel is given by 
\begin{equation}
\kernel{\opr{T}_{2,\gamma}}=
i\hbar\sum_{k,k'}\!' \frac{\phi_{k}^{(\gamma)}\!(q) \bar{\phi}_{k'}^{(\gamma)}(q')}{E_{k}-E_{k'}} \nonumber
\end{equation}
where $\phi_k^{(\gamma)}(q)$ are the energy eigenfunctions and the prime indicates that $k=k'$ is excluded in the summation. The kernel can be shown to be symmetric, so $\opr{T}_{2,\gamma}$ is symmetric. This operator is referred to as the Characteristic Time Operator (CTO) in \cite{gal022}, and its properties for arbitrary Hamiltonian is recently studied in \cite{arai}.  Since the eigenvalues of the Hamiltonian satisfy $\sum_{k,k'}\!\!\!\!\!\!\!\! ' \;\; \left(E_{k}-E_{k'}\right)^{-2}< \infty$, the operator $\opr{T}_{2,\gamma}$ is a Hilbert-Schmidt operator; since $\opr{T}_{2,\gamma}$ is symmetric it is a compact self-adjoint operator. The CTO is, by construction, a PTT solution to the TE-CCR, with canonical domain consisting of all vectors $\psi(q)=\sum_{k=-\infty}^{\infty}\psi_k \phi_k^{(n)}(q)$ in the domain of $\opr{H}_{\gamma}$ such that $\sum_{k=1}^{\infty}\psi_k=0$. No non-zero vector is orthogonal to the canonical domain. Hence, the operator $\opr{T}_{2,\gamma}$ is a time operator of dense category.

Since  $\opr{T}_{2,\gamma}$ is compact, it likewise possesses a complete set of eigenvectors belonging to the system Hilbert space, and a discrete spectrum. The eigenvectors and eigenvalues of $\opr{T}_{2,\gamma}$ are computed in energy representation, where the Hamiltonian $\opr{H}$ is the infinite diagonal matrix $H_{kl}=E_k\delta_{kl}$, and the time operator $\opr{T}_{2,\gamma}$ is the infinite matrix $T_{k,l}=i\hbar(E_k-E_l)^{-1}$ for $k\neq l$ and $T_{k,k}=0$. Due to the compactness of $\opr{T}_{2,\gamma}$, an approximate solution to the eigenvalue problem for $\opr{T}_{2,\gamma}$ is obtained by truncating its infinite matrix representation, followed by solving the eigenvalue problem for the truncated matrix. We find that the eigenvectors $\opr{T}_{2,\gamma}$ assume the form $\psi_{n,\gamma}^{\pm}(q)$, $n=1,2,3,\dots$, where the corresponding eigenvalues are ordered according to 
%$\tau_1^+=-\tau_1^->\tau_2^+=-\tau_2^- >\tau_3^+=-\tau_3^- \dots$. The eigenvalues can be expressed as 
$\tau_{n}^{\pm}=\pm\tau_{n}$, with $\tau_{1}>\tau_{2}>\tau_{3}\dots>0$. 

\subsection{Physical Interpretation of the CTO}

%%%%%%%%%%%%%%%%%%%%%%%%%%%%%%%%%%%%%%%%%%%%%%%%%%%%%%%%%%%%%%%%%%%%%%%%%%%%%%%%%%%%%
%%%%%%%%%%%%%%%%%%%%%%%%%%%%%%%%%%%%%%%%%%%%%%%%%%%%%%%%%%%%%%%%%%%%%%%%%%%%%%%%%%%%%
\begin{figure}
\includegraphics[width=0.23\textwidth,height=0.18\textwidth]{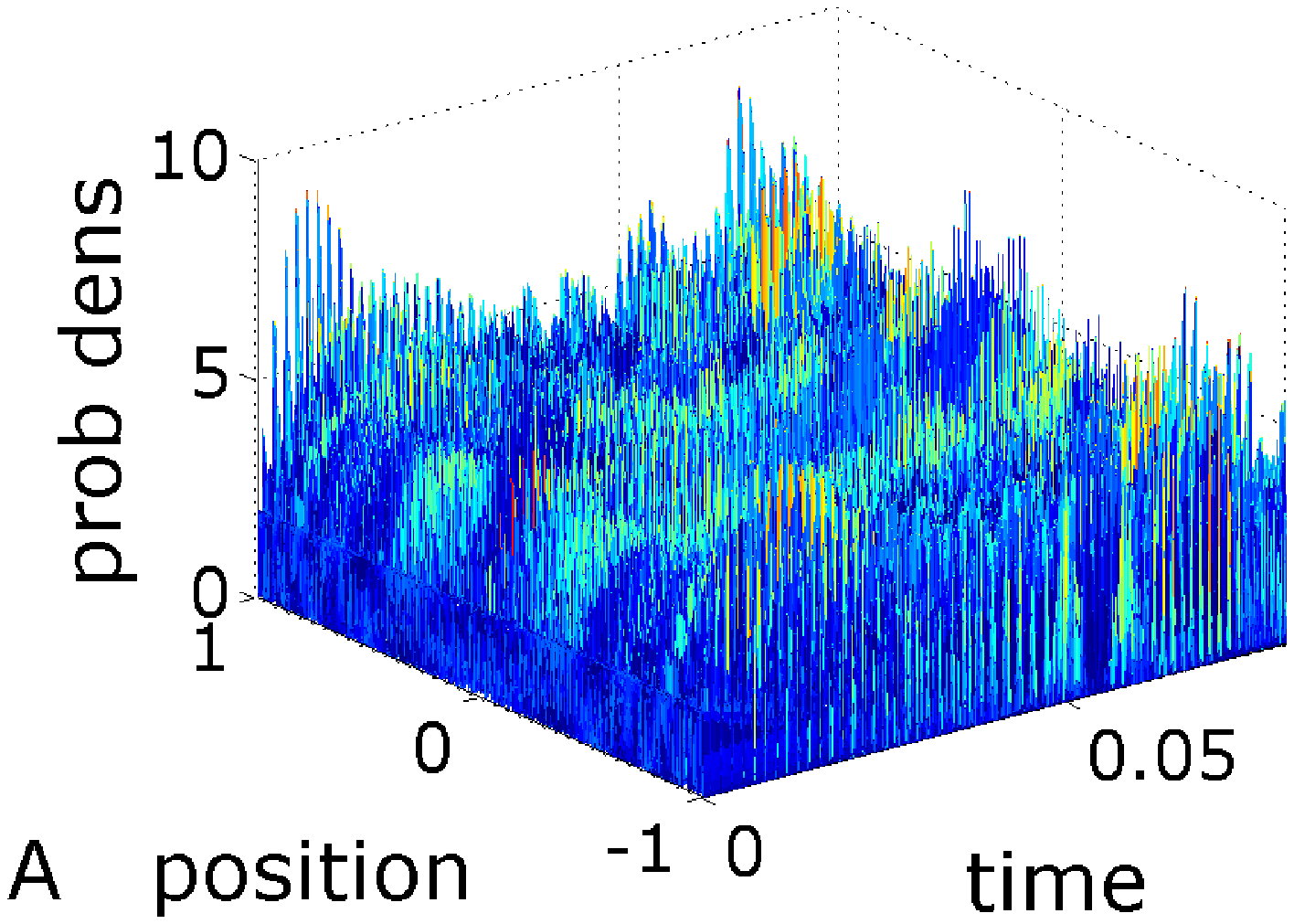}
\includegraphics[width=0.23\textwidth,height=0.18\textwidth]{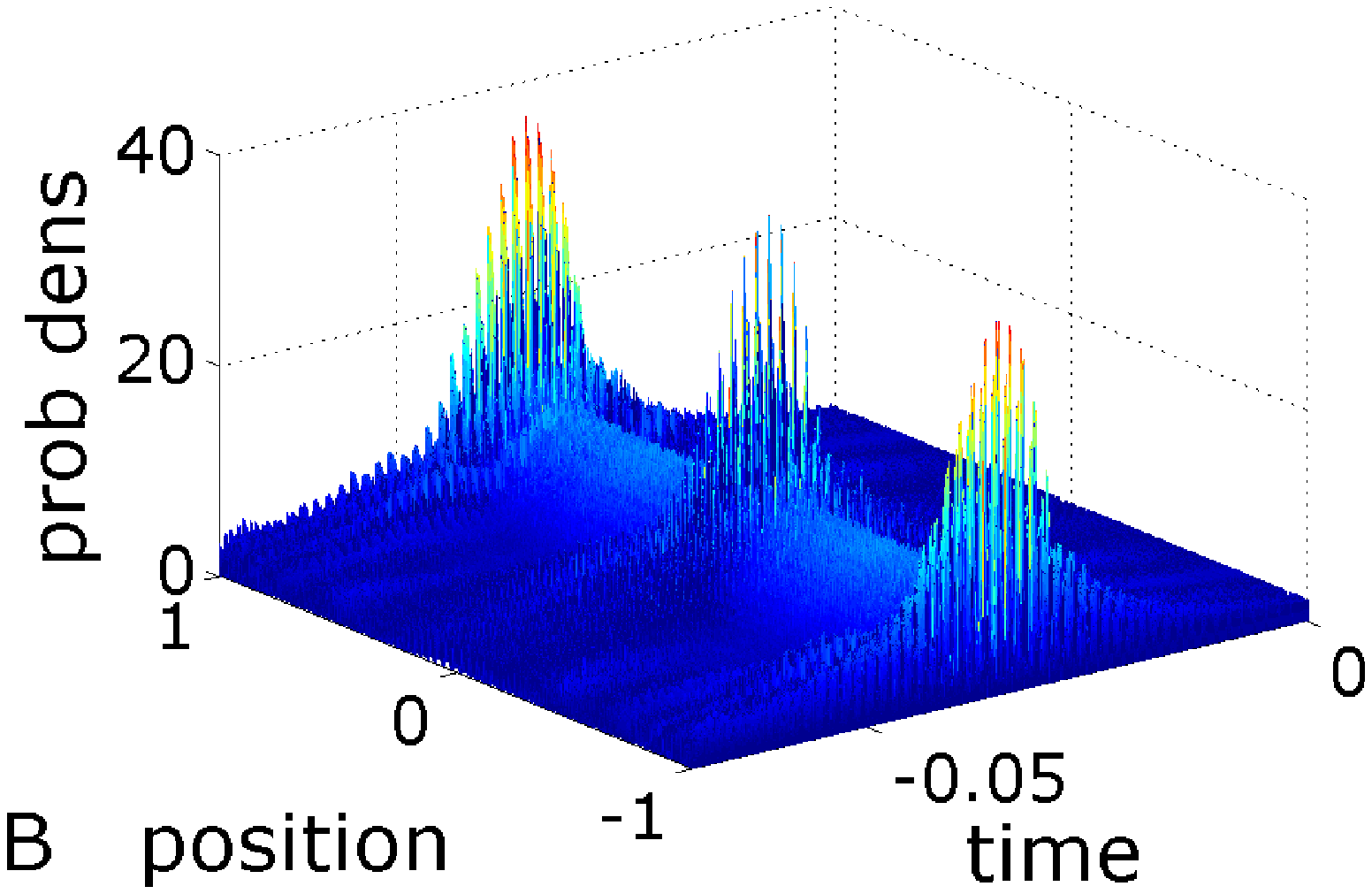}
\includegraphics[width=0.23\textwidth,height=0.18\textwidth]{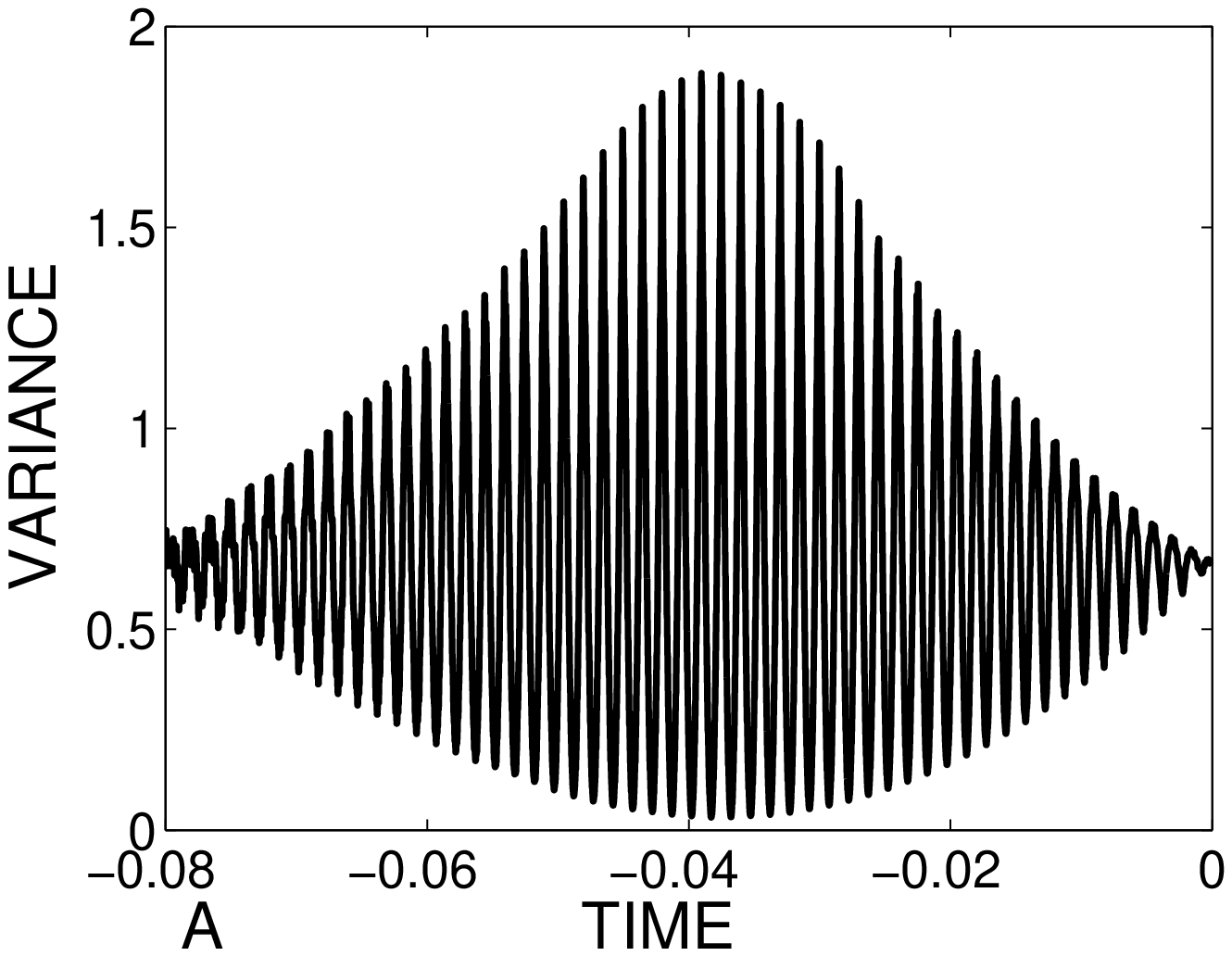}
\caption{(A) Plot of the position probability density of the forward time evolved CTO eigenfunction corresponding to eigenvalue $\tau_{n}=0.03521$. (B) Plot of the position probability density of the backward time evolved CTO eigenfunction corresponding to eigenvalue $\tau_{n}=0.03521$. (C) Plot of the variance of the backward time evolved CTO eigenfunction corresponding to eigenvalue $\tau_{n}=0.03754$. }
\end{figure}
%%%%%%%%%%%%%%%%%%%%%%%%%%%%%%%%%%%%%%%%%%%%%%%%%%%%%%%%%%%%%%%%%%%%%%%%%%%%%%%%%%%%%
%%%%%%%%%%%%%%%%%%%%%%%%%%%%%%%%%%%%%%%%%%%%%%%%%%%%%%%%%%%%%%%%%%%%%%%%%%%%%%%%%%%%%

Now we determine if it is possible to interpret the CTO as a time of arrival operator in the same way as the CTOA-operator is. Evolving the eigenfunctions of $\opr{T}_{2,\gamma}$ forward in time shows no evidence of unitary arrival, as shown in Figure 2.A. But this can be expected because $\opr{T}_{2,\gamma}$ is a PTT time operator not a TAT time operator. But we can define the operator $\opr{T}'_{2,\gamma}=-\opr{T}_{2,\gamma}$, which is, by definition, a TAT time operator. Evolving the eigenfunctions of $\opr{T}'_{2,\gamma}$ forward in time or the eigenfunctions of $\opr{T}_{2,\gamma}$ backward in time shows three peaks converging at three separate neighborhoods. These peaks are themselves spread over a wide time interval. This is illustrated in Figure 2.B. Also, Figure 2.C shows the variance of the backward time evolved eigenfunctions as a function of time, which is very different from the variance plot of the time evolved eigenfunctions of the CTOA eigenfunctions. These show that the eigenfunctions have a drastically different dynamical behavior as those of the CTOA-operators. The CTO-operators then cannot be interpreted in terms of the unitary arrival of its time evolved eigenfunctions at the origin, signifying that they are distinct from the CTOA operator with respect to their dynamics, even if mathematical transformations are performed on them to formally transform them into TAT time operators. It is now evident that one solution to the TE-CCR can be distinguished from another solution.

%%%%%%%%%%%%%%%%%%%%%%%%%%%%%%%%%%%%%%%%%%%%%%%%%%%%%%%%%%%%%%%%%%%%%%%%%%%%%%%%%%%%%%
%%%%%%%%%%%%%%%%%%%%%%%%%%%%%%%%%%%%%%%%%%%%%%%%%%%%%%%%%%%%%%%%%%%%%%%%%%%%%%%%%%%%%%
\begin{figure}
\includegraphics[width=0.23\textwidth,height=0.18\textwidth]{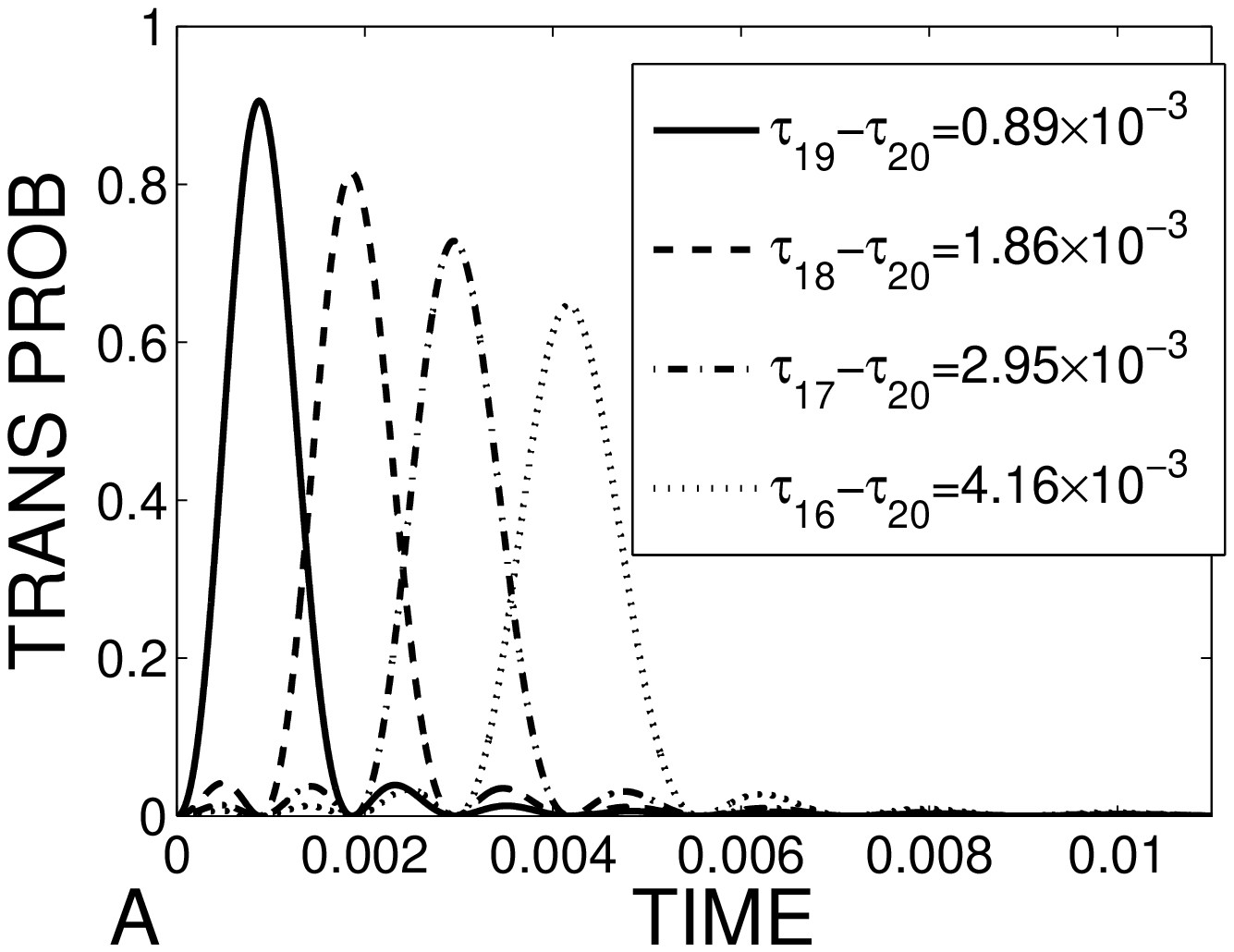}
\includegraphics[width=0.23\textwidth,height=0.18\textwidth]{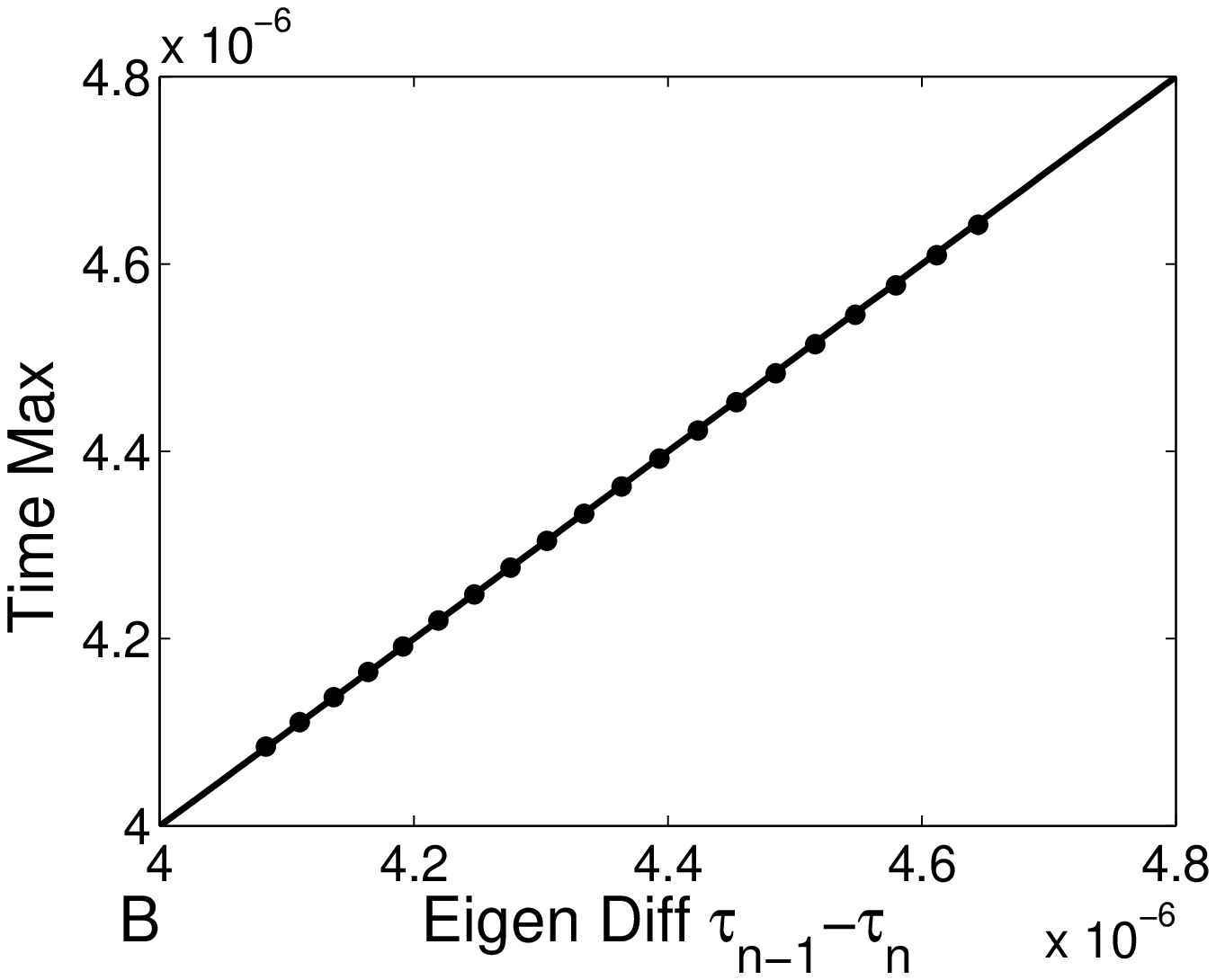}
\includegraphics[width=0.23\textwidth,height=0.18\textwidth]{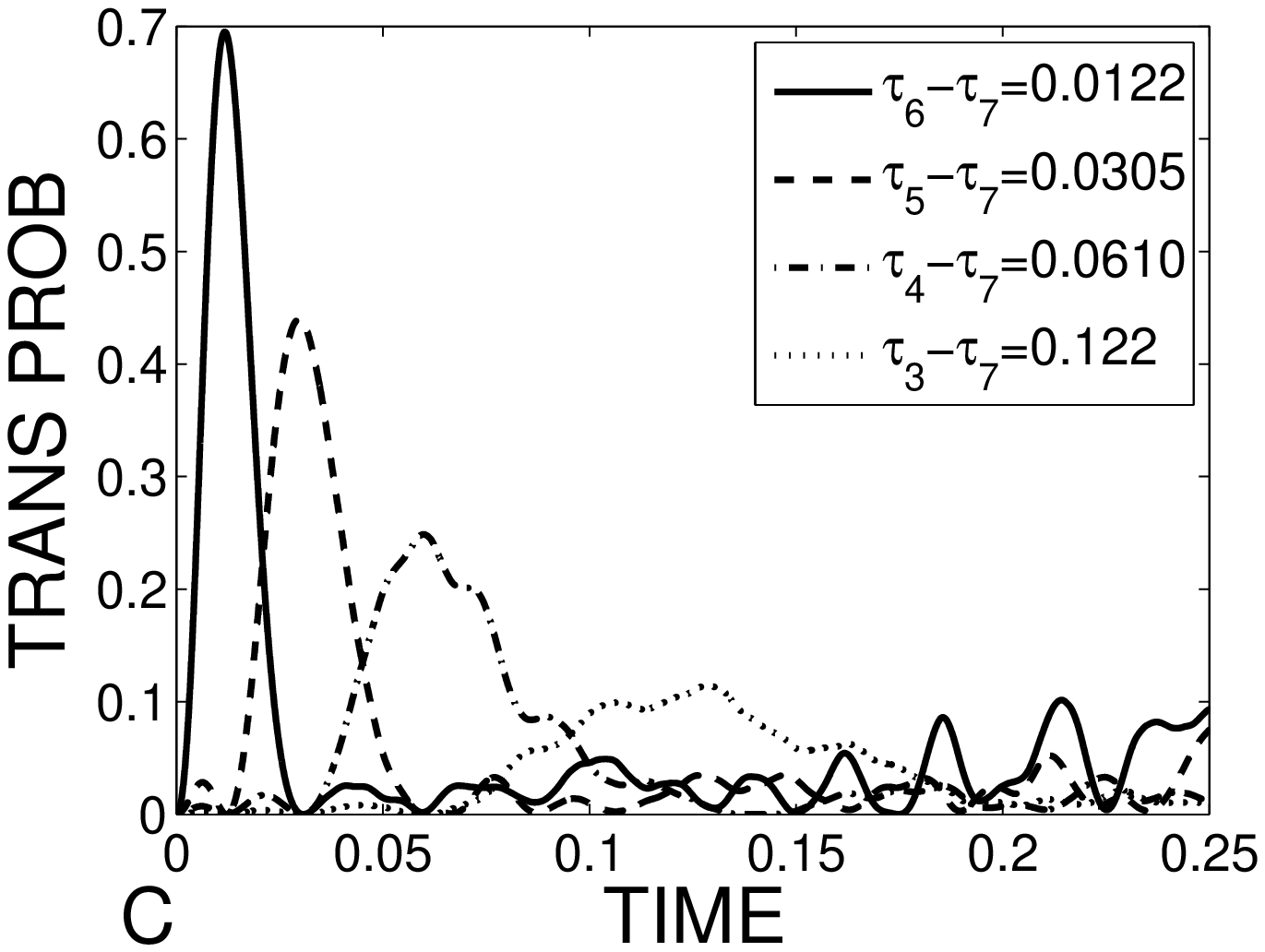}
\includegraphics[width=0.23\textwidth,height=0.18\textwidth]{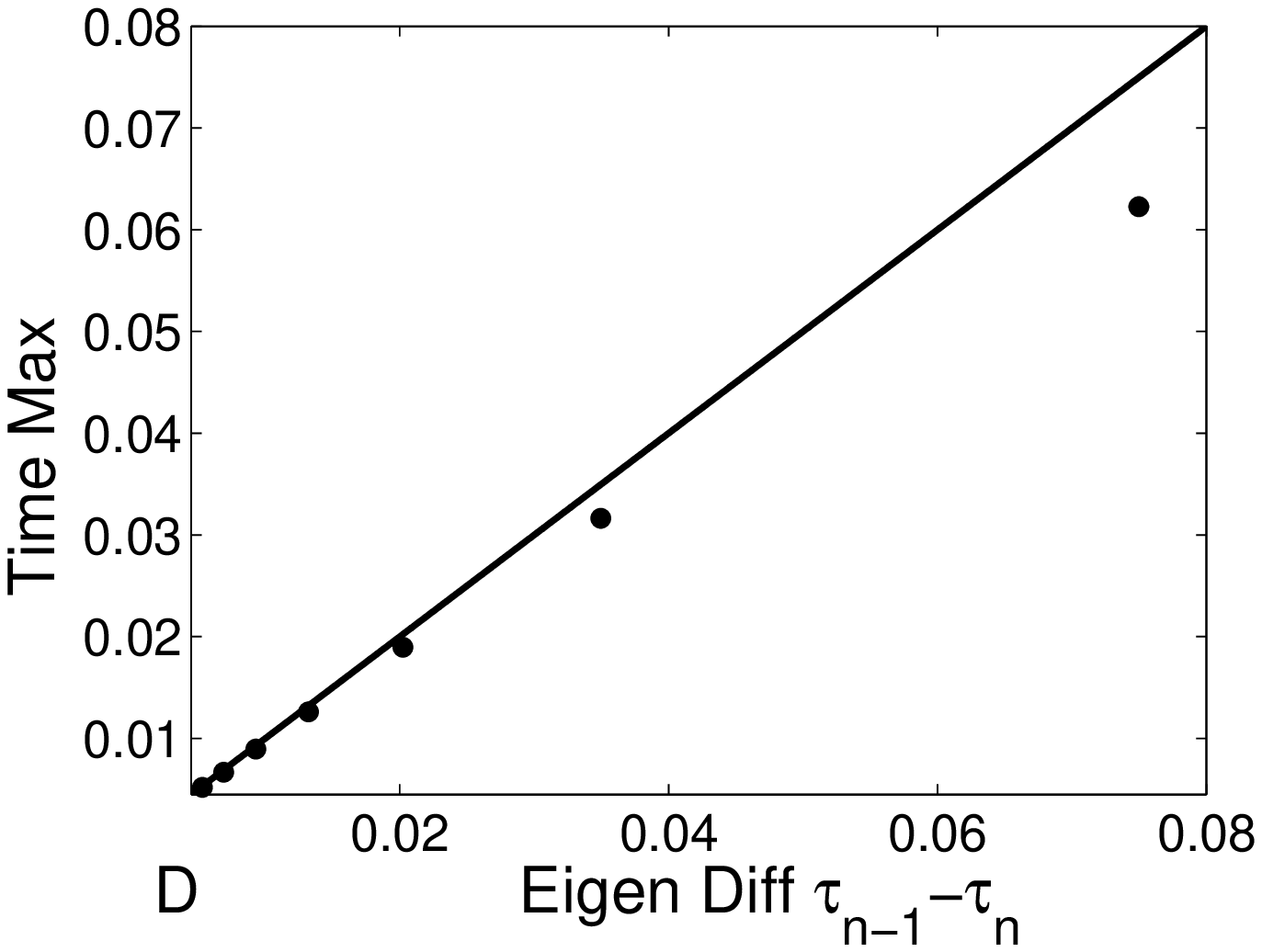}
\caption{ (A and C) Plot of the transition probability $P_{t}[n,n']$ of the CTO eigenfunction pair $\left\{\psi_{n,\pi/6}^{+}(q,t), \psi_{n',\pi/6}^{+}(q)\right\}$, with  eigenvalue differences $\tau_{n'}^{+}-\tau_{n}^{+}$ indicated. (B and D) Plots of the times when the transition probabilities $P_{t}[n,n-1]$ will reach maximum values as functions of $\left(\tau_{n}^{+}-\tau_{n-1}^{+}\right)$, with the range of values of $n$ given for each figure as follows: (\textit{B}) $300\leq n\leq 320$. (\textit{D}) $2\leq n\leq 7$.}
\end{figure}
%%%%%%%%%%%%%%%%%%%%%%%%%%%%%%%%%%%%%%%%%%%%%%%%%%%%%%%%%%%%%%%%%%%%%%%%%%%%%%%%%%%%%%
%%%%%%%%%%%%%%%%%%%%%%%%%%%%%%%%%%%%%%%%%%%%%%%%%%%%%%%%%%%%%%%%%%%%%%%%%%%%%%%%%%%%%%

Let us now examine the transition probability between two CTO eigenvectors, $P_{t}^{+}[n,n']=\left|\left\langle\psi_{n',\gamma}^{+}\right.\left|\opr{U}_{t}\psi_{n,\gamma}^{+}|\right\rangle\right|^2$. 
, where $\opr{U}_{\pm t}$ is the time evolution operator, and the evolution is forward in time. Figure 3 summarizes the results of our analysis. Figures 3.B and 3.D shows that $P_{t}^{+}[n,n']$ will attain a maximum value at an instant of time $t_{max}\approx\tau^{+}_{n'}-\tau^{+}_{n}$ (as indicated by the linear plot with slope 1 in figure 3.B), so long as $\left|\tau_{n'}^{+}\right|>\left|\tau_{n}^{+}\right|$ and $\tau^{+}_{n'}-\tau^{+}_{n}$ is very small (as shown by figure 3.B and the linear plot with slope not equal to 1 in figure 3.D). Furthermore, as shown in figures 3.A and 3.C, if $\tau^{+}_{n'}-\tau^{+}_{n}$ is very small, $P_{t}^{+}[n,n']$ will have a single narrow peak, with the tip of that peak reaching almost to 1, signifying that the transition of the time evolved CTO eigenfunction $\ket{\opr{U}_{t}\psi_{n,\gamma}^{+}}$ to the CTO eigenfunction $\ket{\opr{U}_{t}\psi_{n',\gamma}^{+}}$ will have a high probability of occuring. One would also obtain similar results if one is to examine the transition probability $P_{-t}^{+}[n,n']=\left|\left\langle\psi_{n',\gamma}^{-}\right.\left|\opr{U}_{-t}\psi_{n,\gamma}^{-}\right\rangle\right|^2$, where the evolution is backward in time and the CTO eigenfunctions correspond to negative eigenvalues $\tau_{n'}^{-}=-\tau_{n'}^{+}$ and $\tau_{n}^{-}=-\tau_{n}^{+}$ respectively.

These results can then be used to formulate the following physical interpretation of the CTO: it is a time operator whose time evolved eigenfunctions have a very high probability of making transitions to other CTO eigenfunctions at an instant of time equal to the difference between their corresponding eigenvalues, provided that the eigenvalues have the same sign and that the difference between the corresponding eigenvalues asymptotically approaches zero. This physical interpretation of the CTO implies that one can use the CTO eigenstates as quantum clock states. However, a full investigation into the use of the CTO eigenstates as quantum clock states is beyond the scope of this paper and will be left for future consideration.

\subsection{Difference Between the Transitions of the Time Evolved CTO Eigenfunctions and the Time Evolved CTOA Eigenfunctions}

\begin{figure}
\includegraphics[width=0.23\textwidth,height=0.18\textwidth]{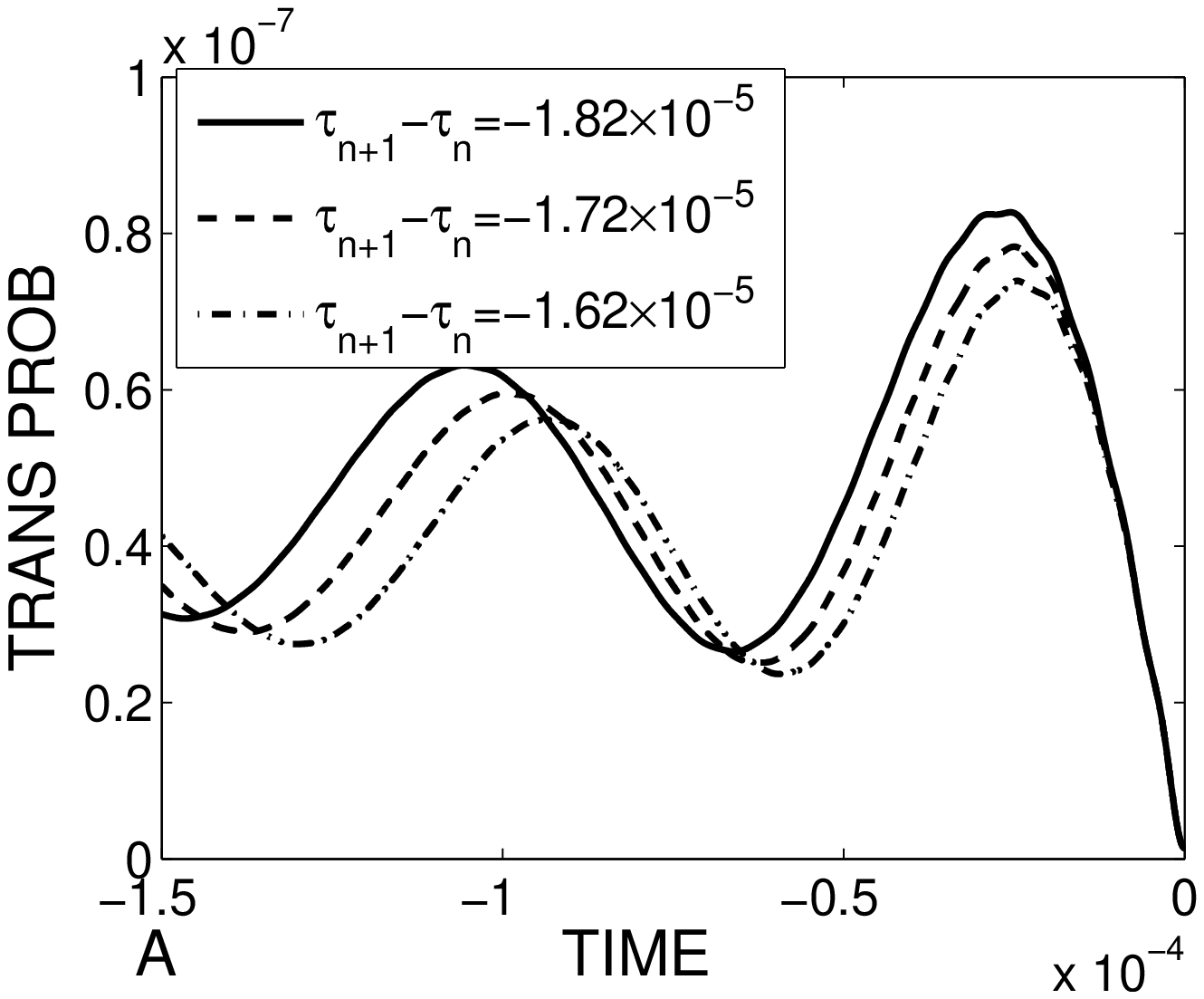}
\includegraphics[width=0.23\textwidth,height=0.18\textwidth]{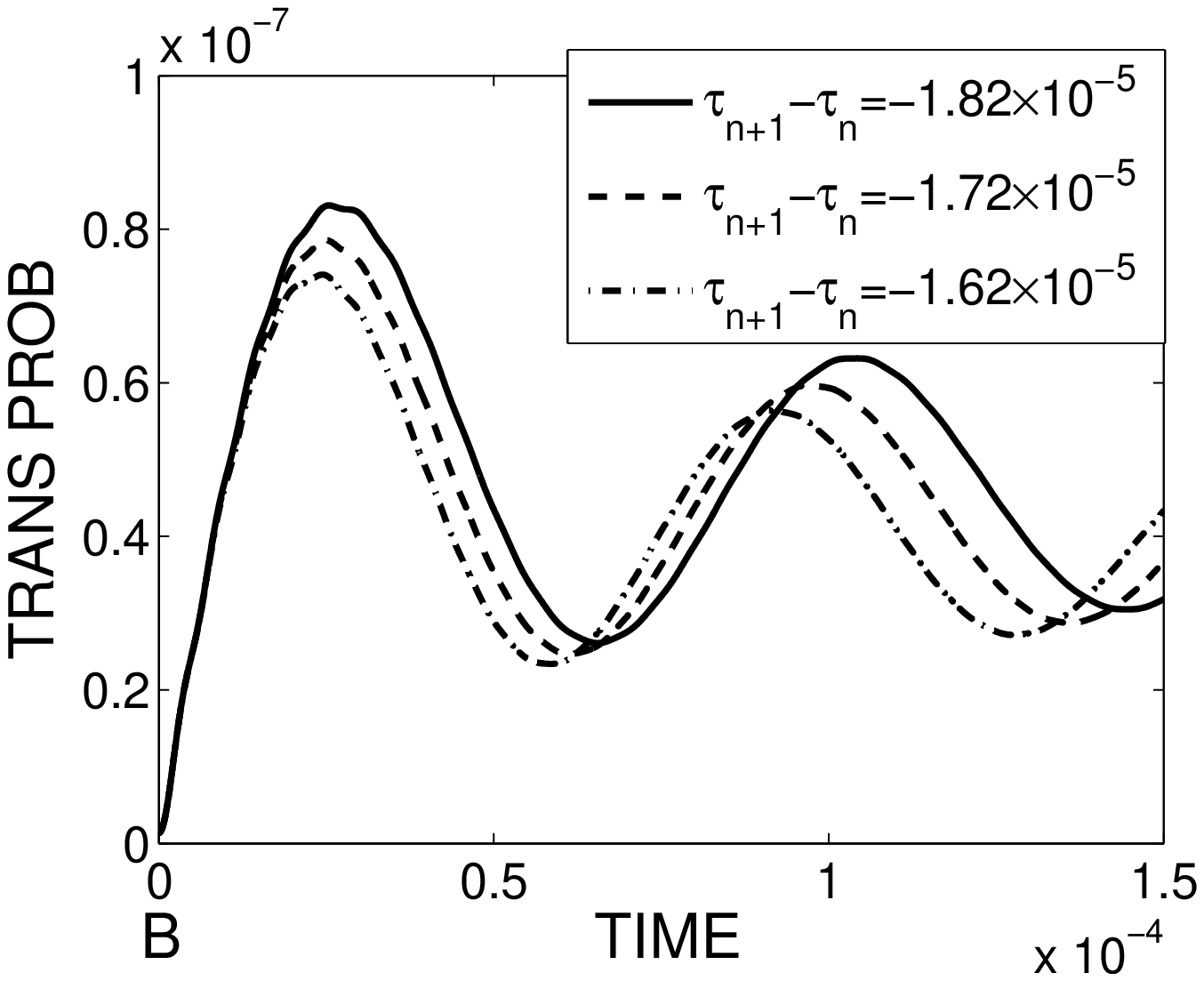}
\includegraphics[width=0.23\textwidth,height=0.18\textwidth]{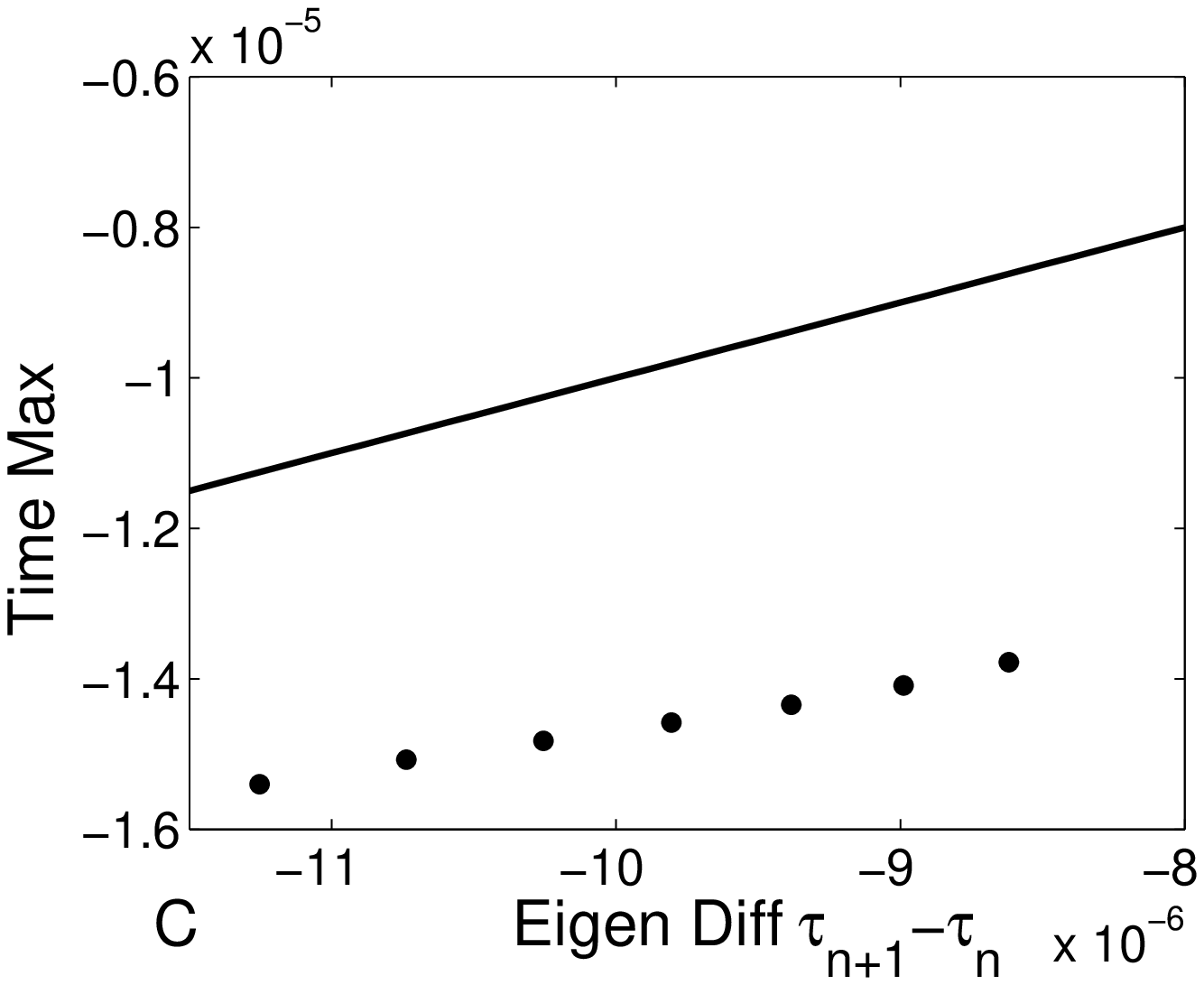}
\caption{Plot of the transition probability $P_{n,n+1}(t)$ of the time evolved CTOA eigenfunction pair $\left\{\varphi_{n,0.01}^{+}(q,t), \varphi_{n+1,0.01}^{+}(q)\right\}$, with $\varphi_{n,0.01}^{+}(q)$ evolved backward (A) and forward (B) in time. The eigenvalue differences are indicated. Note that graph (A) is the mirror image of graph (B). (C) Plot of the times when the transition probabilities of the time evolved CTOA eigenfunctions $P^{+}_{t}[n,n+1]$ will reach maximum values as functions of $\left(\tau_{n+1,\gamma}-\tau_{n,\gamma}\right)$, with $\gamma=0.01$.}
\end{figure}

Let us now address the question of interpreting the CTOA operator in terms of the transition of its time evolved eigenfunctions to other CTOA eigenfunctions. As shown in Figure 4, no matter the direction in which the CTOA eigenfunctions are evolved over time, and whether or not the CTOA eigenfunctions correspond to positive or negative eigenvalues, they will not make transitions to other CTOA eigenfunctions at the instant of time $t=\tau_{n',\gamma}^{\pm}-\tau_{n,\gamma}^{\pm}$ (as evidenced by the plot in figure 4.C). Furthermore, figures 4.A and 4.B signify that no matter the direction of the time evolution, transitions between time evolved CTOA eigenfunctions to other CTOA eigenfunctions will have a small probability of occuring. This implies that transitions by time evolved CTOA eigenfunctions to other CTOA eigenfunctions are distinct from transitions by time evolved CTO eigenfunctions to other CTO eigenfunctions, so the CTOA operator cannot be interpreted in terms of the transition probabilities of its time evolved eigenfunctions, thus directly answering the question we raised at the end of section 4 of this paper. This further implies that even if the sign of the CTOA operator $\opr{T}_{1,\gamma}$ were changed (that is, if we define an operator $\opr{T}'_{1,\gamma}=-\opr{T}_{1,\gamma}$), thus formally transforming the TAT type operator $\opr{T}_{1,\gamma}$ to a PTT type operator, the time evolved eigenfunctions of $\opr{T}'_{1,\gamma}$ will still not make transitions to other eigenfunctions of $\opr{T}'_{1,\gamma}$ at an instant of time equal to the difference between their corresponding eigenvalues. Hence, with respect to their dynamics, the CTOA operator is distinct from the CTO, even if mathematical transformations are performed, formally transforming the CTOA operator into a PTT time operator. 

\section{Intrinsic PTT and TAT Time Operators}

The results presented in the preceding sections show that the dynamical behavior of the CTOA operator and the CTO are distinct from each other and correspond to the expected dynamical behavior for a TAT and PTT time operator, since the dynamics of a TAT time operator correspond to some form of arrival at a given instant of time while the dynamics of a PTT time operator corresponds to the passage of a given interval of time. We have also seen that even if the sign of the CTOA operator and CTO were changed, the dynamical behavior of the resulting operator is not the same as the dynamical behavior of the CTOA and the CTO, respectively. These findings then signify that the CTOA operator and the CTO are intrinsically TAT and PTT time operators, respectively.

These results imply that changing the sign of a TAT or PTT time operator will not transform it into a PTT or TAT time operator, contrary to what one expects when, based on the explicit forms of the TE-CCR that TAT and PTT time operators obey, one changes the sign of a TAT or PTT time operator. This also implies that the seemingly trivial mathematical difference between TAT and PTT time operators due to the difference in sign of the TE-CCR which they obey is not so trivial after all, since the mathematical difference in sign masks a fundamental physical difference between TAT and PTT time operators.

\section{Conclusion}

We have shown in this paper that it is possible to differentiate between multiple solutions to the TE-CCR, and at the same time provide unique physical interpretations of these solutions to the TE-CCR, using the internal unitary dynamics of the system, thus directly answering in the affirmative the question of whether there are identifying differences between solutions belonging to different categories, which we raised in section 2. This is significant since the work shows that multiple solutions to the TE-CCR can exist, and each one is physically different from the others. Previous works on the TE-CCR have focused more on determining the mathematical form of the solution to the TE-CCR given the Hamiltonian corresponding to the system via quantization of a known classical time observable without regard to the physical consequences or physical interpretation of the existence of such a quantum observable \cite{blanchard, allcock, kijowski, busch1, muga, muga2, pegg, grot, anas}. In these works, what is only being considered is whether the solution to the TE-CCR is self - adjoint and canonically conjugate to the Hamiltonian, or canonically conjugate to the Hamiltonian and covariant. On the other hand, there are treatments of time as a quantum mechanical observable which focuses less on the observable satisfying the TE-CCR and more on the physical content of the quantum time observable by considering the physical system in which the quantum time observable is defined, and constructing a quantum time observable for that system in terms of other quantum time observables, such as energy \cite{buttiker, bsm, sokolovski, sokolovski2, munoz, halliwell, yamada, hartle, misra, schulman}. This paper then provides a bridge between the two approaches to the treatment of time as a quantum mechanical observable: it shows that one can determine a quantum time observable by first determining whether it is a solution to the TE-CCR, then using the system's internal unitary dynamics to determine the physical significance and physical content of that observable. 

We have also shown in this paper that there are time operators that are intrinsically PTT or TAT time operators by virtue of their dynamical behavior, and we have also shown that seemingly trivial mathematical differences between PTT and TAT time operators actually disguise profound physical differences between these two types of time operators. This finding is significant since it shows that one cannot just construct a quantum time operator and call it either a PTT or TAT time operator simply by quanitizing a known classical time observable or by simply solving the TE-CCR or even by simply transforming one type of time operator mathematically into the other. As shown in this paper, one has to determine the dynamical behavior of a given time operator in order to properly classify it as either a PTT or TAT time operator.  

Furthermore, this work shows that different solutions to the TE-CCR correspond to different aspects of time in quantum mechanics. For instance, the CTOA operator is appropriate for describing the arrival time of quantum particles, and the CTO is appropriate for describing quantum clock states which mark off time via quantum transitions, due to the dynamics of their corresponding eigenfunctions. Generalizing these results, we then say that multiple solutions to the TE-CCR are reflective of the multifaceted nature of time. A solution to the TE-CCR cannot then simply be considered a time operator; one has to consider in what sense is it a time operator. As was shown in this work, both the CTO and the CTOA operators are time operators in the sense that both satisfy the TE-CCR; however, both are different from each other in the sense that they have different physical interpretations with respect to the system's internal unitary dynamics. This then implies that one cannot arbitrarily assign a physical interpretation to a given solution to the TE-CCR, no matter the physical system in which it is constructed. 

Most importantly, however, these results hint at a heretofore unsuspected link between time as a quantum mechanical observable and the internal unitary dynamics of the system. This link arises as a consequence of solutions to the TE-CCR being interpreted in terms of the system's internal unitary dynamics. The exact nature of that relationship can only be made clear if we undertake a closer look not just at the role of time in quantum mechanics, but at the foundations of quantum mechanics itself.

\textbf{Acknowledgment}
This work has been supported by UP System Grant.

\end{document}